\documentclass[journal]{IEEEtran}
\usepackage{times,url}
\usepackage{graphics,graphicx,setspace}
\usepackage{amssymb,amsmath,amsbsy, amstext, amsthm}
\usepackage[ruled,linesnumbered]{algorithm2e}

\usepackage{float}
\usepackage{subfigure}

\newtheorem{lemma0}{\bf Lemma}

\newtheorem{definition0}{\bf Definition}

\begin{document}
%
\title{Uplink Scheduling Strategy Based on A Population Game in Vehicular Sensor Networks}
\author{\IEEEauthorblockN{Jiajun Sun
}

\IEEEauthorblockA{Beijing Key Laboratory of Intelligent
Telecommunications Software and Multimedia,\\
Beijing University of Posts and Telecommunications,\\
Beijing 100876, China\\
Email: jiajunsun@bupt.edu.cn}
}


\maketitle

\begin{abstract}
Recent advances in the integration of vehicular sensor network (VSN)
technology, and crowd sensing leveraging pervasive sensors called
onboard units (OBUs), like smartphones and radio frequency
IDentifications to provide sensing services, have attracted
increasing attention from both industry and academy. Nowadays,
existing vehicular sensing applications lack good mechanisms to
improve the maximum achievable throughput and minimizing service
time of participating sensing OBUs in vehicular sensor networks. To
fill these gaps, in this paper, first, we introduce real imperfect
link states to the calculation of Markov chains. Second, we
incorporate the result of different link states for multiple types
of vehicles with the calculations of uplink throughput and service
time. Third, in order to accurately calculate the service time of an
OBU, we introduce the steady state probability to calculate the
exact time of a duration for back-off decrement, rather than using
the traditional relative probability. Additionally, to our best
knowledge, we first explore a multichannel scheduling strategy of
uplink data access in a single roadside unit (RSU) by using a
non-cooperative game in a RSU coverage region to maximize the uplink
throughput and minimize service time under saturated and unsaturated
traffic loads. To this end, we conduct a theoretical analysis and
find the equilibrium point of the scheduling. The numerical results
show that the solution of the equilibrium points are consistent with
optimization problems.
\end{abstract}

\IEEEpeerreviewmaketitle

\section{Introduction}\label{sec:intro}
Vehicular Sensor Networks connected to the Internet backbone or
various other application servers is emerging as a new network
paradigm for sensed information sharing in urban environments
\cite{wang20113g}. With the advent of 4G networks and more powerful
processors, smartphones have received a lot of attention for their
potential as portable vehicular urban sensing platforms, as they are
equipped with a variety of environment and motion sensors (e.g.,
audio/video, accelerometer, and GPS) and multiple wireless
interfaces (e.g., WiFi, Bluetooth and 2/3G). The ability to take a
smartphone on board to complement the sensors of the latter with
advanced smartphone capabilities is of immense interest to the
industry \cite{lee2010survey}. Recent these advances make it
possible for most of OBUs on the road to use vehicle-to-roadside
(V2R) communication \cite{bruno2011performance}. As such, VSN can
support a wide range of applications for improving the road
transport efficiency. In V2R communication, all vehicles within the
coverage region of a roadside unit (RSU) should be associated with
the RSU, which is responsible for all the communications between the
vehicles, such as broadcasting of control and safety messages,
inter-vehicle data transfer, non-safety message transmission, etc.
Thus, in vehicular networks, V2R communications are the preferred
way for OBUs on the roadway to a
wide range of applications. 

Recently, many researchers has focused on performance analysis of
V2R communications, since these RSUs like 802.11(WiFi) can provide
data transfers of broadband speeds for OBUs in its coverage region
including highly mobile users traveling in cars
\cite{eriksson2008cabernet}. Currently, several works have
experimentally validated the feasibility of using 802.11(WiFi) based
RSUs
communications at vehicular speeds. 

As is well known, in static single-hop wireless LAN, the network
performance under saturation and non-saturated assumption has been
greatly enhanced
\cite{bianchi2000performance,kumar2007new,felemban2011single,zhao2011modeling,garetto2005performance,rathod2009characterizing,panda2009state}.
However, in the high speed vehicle networks, this mobility greatly
increases the collisions between simultaneous transmissions of
vehicles contending for the access to the same RSU, resulting in
significant deterioration of the performance of V2R communications.
Therefore, how to assure the reliability and QoS of safety and
non-safety services in this type of V2R communications is a common
concern. To this end, some researchers start to focus on IEEE 802.11
DCF scheme in V2R communications
\cite{bruno2011throughput,bruno2011performance,misic2011performance,luan2011mac,ge2010optimal}.
From these existing works, we know that most of them focus on a very
sparse OBUs in the coverage region of a single RSU. However, in the
high-speed mobile dense scenario, V2R communications utilizing
802.11 DCF scheme face the following new challenges as follows:

\begin{itemize}
\item The inaccuracy in modeling the back-off process about 802.11 DCF scheme is
one of the main reasons for deterioration of the communication
performance in V2R communications. If we are able to highly
accurately calculate the back-off time by some models, the data
transfer can be reasonably scheduled, thereby, the performance in
V2R communications will be greatly enhanced.
\item Real traffic flows are abrupt and do not
generate true saturation in V2R communications. OBUs with the
transient empty queue are also counted in the scheduling algorithm.
This is an urgent need for an adaptive algorithm to solve this real
unsaturated flow impact on the performance.
\item This single-channel data transfers are difficult to adapt to
the growing demands in V2R communications.
\end{itemize}

With the emergence of low-cost 802.11-based WiFi devices and the
advances of multichannel wireless technology, it is reasonable to
expect to enjoy data transfers at broadband speeds by connecting to
multichannel wireless devices
\cite{eriksson2008cabernet,aryafar2012midu,tan2009sam,aryafar2010design}.
Thus, each RSU can provide the Internet access to dense vehicles
simultaneously and efficiently.

In addition, to meet demands in V2R communications, the U.S Federal
Communication Commission (FCC) allocates 75MHz of spectrum ranges
between 5.850 to 5.925 GHz band for VSN. Further, the 75MHz spectrum
ranges are partitioned into seven non-overlapping channels by
category, one for safety application control channel (CCH) and the
others for service channel (SCH) providing non-safety data
transmissions such as Internet services and video on demand run on.
Further, the authors of \cite{aryafar2012midu} provides guidelines
for the design of an efficient MAC for single cells employing MIDU
nodes. More importantly, it scales very easily to MIMO systems and
provides large self-interference cancelation no matter transmission
or reception is performed simultaneously, thereby make the
co-existence of MIMO with full duplex possible. Recent works
\cite{tan2009sam,aryafar2010design}, have implemented Multi-User
MIMO schemes, in which an RSU can communicate with a number of OBUs
simultaneously by utilizing the antennas that belong to a group of
OBUs.

Therefore, in this paper, we mainly focus on situations where the
overall loads generated by high density vehicles are too heavy,
i.e., packet collisions are too many, thus many traffic applications
cannot be supported satisfactorily. To avoid such problems, we need
to provide an alternative mechanism which is able to guarantee
maximizing their individual throughput and minimizing the service
time. Specifically speaking, our main results and contributions are
summarized as follows:
\begin{itemize}
\item To our best knowledge, we are the first attempt to
explore a multichannel scheduling strategy of uplink data access in
a single RSU by using a non-cooperative game in V2R communications
to maximize the uplink throughput and minimize service time under
saturated and unsaturated traffic loads. To this end, we conduct a
theoretical analysis to find the balance point of the scheduling.
The numerical results show that the solution of the equilibrium
points are consistent with optimization problems.
\item In order to accurately calculate the service time of an OBU, we
introduce the steady state probability to calculate the exact time
of a duration for back-off decrement, rather than using the
traditional relative probability. Further, we explore saturated and
unsaturated traffic loads to accurately estimate the MAC-layer
uplink throughput and service time by a calculation of back-off
freezing probability for an arbitrary buffer size under multichannel
conditions.
\item We first introduce real imperfect link
states to the calculation of our vehicle model for Markov chains.
Eventually, the whole system throughput and the service time are
accurately calculated.
\item We incorporate the result of different
link states for multiple types of vehicles with the calculations of
uplink throughput and service time under multichannel conditions in
a dense traffic scenario.
\end{itemize}
%
The rest of the paper is organized as follows. Section \ref{related}
briefly discusses the related work. In Section
\ref{SystemModelDefinition}, we present our system model and related
definitions. We briefly discuss concepts of finite non-cooperative
games and population games. In Section IV, we first analyze the the
actual link state for V2R communications, and then present the
mathematical development of a single type of vehicles model and
multiple types of vehicles model. Second, we incorporate the actual
link state and  multiple types of vehicles model with IEEE 802.11p
Markov chain based on different regions. Additionally, we develop
the expressions of throughput and service time by using the conflict
probability. Section \ref{AnalyticalModel}, we form a
non-cooperative game problem, and then we conduct a theoretical
analysis,and find the balance point of the scheduling. In Section
\ref{MultichannelPopulationGame}, we study the dynamics of the
system in a non-cooperative scenario. The idea here is to show that
the system is stable using Lyapunov techniques. We next study the
efficiency of such an equilibrium and show that the Wardrop
equilibrium is efficient. In Section \ref{CaseStudy}, we make a case
study. The numerical results show that the solution of the
equilibrium points are consistent with optimization problems.
Finally, Section \ref{conclude} presents concluding remarks.

\section{Background and Related Work}~\label{related}
The popular IEEE 802.11 wireless LAN  using a CSMA/CA mechanism
called the Distributed Coordination Function (DCF) is studied
extensively in the literature. The authors of
\cite{bianchi2000performance} focus on obtaining the system
throughput and average long term metrics such as saturation
throughput by using a bi-dimensional discrete-time Markov-chain
model, while two important features specified by IEEE 802.11b
standard, which are retransmission limits and back-off counter
freezing, are not taken into account. The authors of
\cite{kumar2007new} used a renewal theory to develop a fixed-point
formulation relating the per-station attempt rate with the collision
probability of a packet under saturation.

Compared with the traditional decoupling saturation assumption,
non-saturated models differ in approach and scope, but they are all
in some way derived from a saturated fixed-point formulation. The
authors of \cite{garetto2005performance} modified a saturated
fixed-point formulation to overcome these difficulties under
non-saturated conditions. The authors in
\cite{garetto2005performance} analyzed a three-dimensional extension
of the Bianchi Markov chain that explicitly tracked the buffer state
of a station, as well as the number of other stations with a
nonempty buffer. The authors in
\cite{garetto2005performance,ozdemir2006performance} dispense with
the decoupling assumption for the collision probability. The authors
in \cite{rathod2009characterizing} use a three-way fixed point to
model the node behavior with Bernoulli packet arrivals and determine
closed form expressions for the distribution of the time spent
between two successful transmissions in an isolated network. The
authors of \cite{zhao2011modeling} present an accurate non-saturated
model on the saturated renewal process of \cite{kumar2007new} and
extend the buffer size to an arbitrary value for the non-saturated
attempt rate. However, they are only confined to using the relative
probability to calculate the freezing time of a duration for
back-off decrement, which leads to inaccurate calculation of the
frame service time. The authors of \cite{felemban2011single} apply
the idea of the steady probability to calculate transmission
probability for unsaturated traffic cases, while they do not take
the exact calculation of a duration for back-off decrement into
account.

 In V2R communications, the authors of
\cite{bruno2011throughput} are the first to introduce the
traditional decoupling saturation assumption to analyze the maximum
achievable throughput when multiple vehicles simultaneously share
the bandwidth of the same in a given mobility scenario. The authors
of \cite{misic2011performance} are the first to model time division
between CCH and SCH with multiple traffic combinations/classes in
the non-saturation regime under a single channel, assuming that an
OBU's buffer has infinite length and use $M/G/1$ queuing model. The
authors of \cite{luan2011mac} take high node mobilities into account
by using a three-dimensional Markov chain, where the spatial zone is
simple and increase the computational complexity. However, they do
not take the impact of link-state sending rate into account. In
fact, the wireless link states can indeed vary with their locations
and environmental dynamism.

\section{System Model and Related Definition}~\label{SystemModelDefinition}
\subsection{System Model}
In this work we take different type OBUs in the coverage region for
a single RSU into account, which operate uplink data access under
multichannel conditions whether the nodes are saturated or not.
These OBUs are deployed on a road segment. Assuming that there are N
link states, corresponding to the $N$ non-overlapping regions by
thresholds $\Gamma _{f}(f\in\{1, \cdots, N\})$. In each region $f$
within the RSU coverage region, OBUs have different link qualities
resulting in different transmission rates according to the
signal-to-noise ratio (SNR) at the receiver to RSU.
Fig.~\ref{fig:mrsu2} shows our system model in detail. For
analytical convenience, we make the following approximations:

\begin{itemize}
\item Synchronization delays for the OBUs and the
RSU can not occur. Time is slotted with slot length and the back-off
process will be completed in the vicinity of a single RSU. Different
channels are orthogonal and non-interfering.
\item Different types of vehicles (cars, trucks, buses, etc.) can occur in
a single RSU coverage region. all vehicles have the same speed
mobile model and different speed parameters.
\end{itemize}

\subsection{A Population Game Theory}~\label{populationgame}
A population game with $C$ continuous populations is defined by a
mass and a strategy set for each population class and a payoff
function for each strategy, where the set of population classes
$\mathcal{C}=\{1, \cdots, C\}$, each of which corresponds to the
same type of OBUs with the same channel conditions $\{L_{i},
c^{1}_{i}, \cdots, c^{L}_{i}\}$, where these OBUs can choose a
channel from the same channel set $\{c^{1}_{i}, \cdots,
c^{L}_{i}\}$.
and the set of strategies
corresponds to the set of $L$ independent channels which are
orthogonal and so do not interfere with each other, $\mathcal
{S}=\{1, \cdots, L\}$.
Strategies of these (population $i$) OBUs
lead to a strategy distribution $\mathcal {X} _{i}=\{x_{i}\in
R_{+}^{L}: \sum _{j\in \mathcal {S}}x^{j}_{i}=n_{i}\}$. The $n_{i}$
denotes the number of OBUs belonging to the i-th type. As such, the
overall strategy distributions is denoted as $\mathcal
{X}=\{\textbf{x}=(x_{1}, \cdots, x_{C}): x_{i}\in \mathcal {X}
_{i}\}$. 

\begin{definition0}[\textbf{Potential Game}]\label{df:potentialgame}
A potential game holds: There exists a $C^{1}$ potential function of
the game $f : \mathcal {X}\rightarrow R$ such
that$\frac{\partial}{\partial x^{l}_{c}}(f)=F^{l}_{c}(x)$ for all
$x\in\mathcal {X}$, $l\in \mathcal {S}$, and $c\in \mathcal {C}$
\cite{sandholm2001potential}, where $f$ is a continuously
differentiable function which is unique up to an additive constant,
and $F$ is the payoff vector equaling $f$'s gradient.
\end{definition0}
\begin{definition0}[\textbf{Nash Equilibrium}]\label{df:nashequilibrium}
A Nash equilibrium is a state whose support consists solely of best
responses to itself. At a Nash equilibrium, no OBU can unilaterally
improve his payoffs.
\end{definition0}

\begin{definition0}[\textbf{Wardrop Equilibrium}]\label{df:wardropequilibrium}
A state $\hat{x}$ is a Wardrop equilibrium if $\hat{\mathcal
{S}}\subset \mathcal {S}, F^{l}_{c}(\hat{x}) \geq
F^{l'}_{c}(\hat{x}), \forall l\in \hat{\mathcal {S}}$ and $l'\in
\mathcal {S}$ \cite{sandholm2001potential}.
\end{definition0}

\begin{definition0}[\textbf{Positive Correlation}]\label{df:positivecorrelation}
The dynamics $\dot{x}=V(x)$ is called positive correlation(PC) if
$V(x)\cdot F(x)=\sum_{i\in \mathcal{C}}\sum_{j\in \mathcal
{S}}V_{i}^{j}(x)F_{i}^{j}(x)>0, V(x)\neq0$
\cite{sandholm2001potential} \cite{shakkottai2006case}.
\end{definition0}

\begin{lemma0}\label{lm:Lyapunov}
If $F$ is a potential game and V satisfies PC, then the potential
function of $F$ is a global Lyapunov function and all Wardrop
equilibria of $F$ are the stationary points for $\dot{x}=V(x)$.
\end{lemma0}

\begin{lemma0}\label{lm:asymptoticalpoints}
A potential game F, with dynamics V(x) that are PC, has
asymptotically stable stationary points \cite{shakkottai2006case}.
\end{lemma0}

According to lemma \ref{lm:Lyapunov} and lemma
\ref{lm:asymptoticalpoints}, the dynamics $\dot{x}=V(x)$ would
converge to either a Wardrop equilibrium or a boundary point of the
set $\mathbb{B}$\cite{hofbauer2009brown}.

\begin{definition0}[\textbf{Brown-von Neumann-Nash Dynamics}]\label{df:bnn}
The Brown¨Cvon Neumann¨CNash(BNN) dynamics is defined
\cite{hofbauer2009brown} as

\begin{equation}\label{eq:BNN}
\dot{x}_{c}^{l}=V(\textbf{x})=n_{c}k_{c}^{l}-x_{c}^{l}\sum _{c\in
\mathcal {S}}k_{c}^{l},
\end{equation}
where $k_{c}^{l}=max\{F_{c}^{l}(x)-\frac{1}{n^{c}}\sum _{j\in
\mathcal {S}}x_{c}^{j}F_{c}^{j}(x),0\}$.
\end{definition0}

\begin{lemma0}\label{lm:bnnpc}
The system with BNN dynamics satisfies PC. The complete proof that
BNN dynamics are PC is present in \cite{sandholm2001potential}.
\end{lemma0}

As such, if each class follows BNN dynamics, the system is PC
according to lemma \ref{lm:bnnpc}. The detailed proof can be seen in
\cite{shakkottai2006case}.
\section{Analytical Model}~\label{AnalyticalModel}
In this section, we formalize present and develop relative results
used in Section \ref{MultichannelPopulationGame}.
\subsection{Link State Model}
The wireless links between a RSU and an OBU is modeled as a
finite-state Markov chain (FSMC). In a region $f$, we assume that
$c_{f}$ packets can be transmitted in one transmission period.
Assume that $\hat{\mathbb{C}}^{c_{f}}$ is the diagonal probability
matrix corresponding to the transmission of $c$ packets from a RSU
to an OBU at location $L_{l}$ of the region $f$. The elements of
this matrix are denoted as $\hat{\textbf{C}}_{l}^{c_{f}}(f,f)$,
which is defined as the probability that $c_{f}$ packets are
successfully transmitted when the channel changes from state $f$ in
the current transmission period to state $f$ in the following
transmission period, i.e., the OBU is in the region $f$, can be
given by \cite{niyato2010unified} as follows:

\begin{equation}\label{avergeacounteq}
\hat{\textbf{C}}_{l}^{c_{f}}(f,f)=\left\{ {\begin{array}{*{20}c}
   {\sum\limits_{c = c^f }^{c^f u} {\left( {\begin{array}{*{20}c}
   c\\
   {c^f }\\\end{array}}\right)}(P_s)^{c^f} (1-P_s)^{c-c^f}} & {u\geq 1~~~~~~~~}\\
   0~~~~~~~~~~~~~~~~~~~~~~~~~~~~~~~~~~~ & {otherwise~~~~~~~~~~}\\
\end{array}} \right.
\end{equation}
where $u\in\{0,1,\cdots,R_{max}\}$ is the amount of rate demand for
real wireless transmission and $P_{s}$ is defined in Section
\ref{qwertasdf}. $R_{max}$ is the maximum rate from the RSU in the
region $f$.

\begin{figure}
\setlength{\abovecaptionskip}{0pt}
\setlength{\belowcaptionskip}{10pt} \centering
\centering
\includegraphics[width=0.32\textwidth]{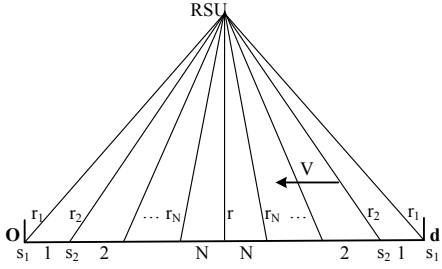}
\caption{The optimal channel game problem for different link
states.} \label{fig:mrsu2}
\end{figure}

From \cite{liu2004cross,niyato2010unified,iskander2003fast}, the
average packet transfer rate from an OBU at region $f$ to the RSU,
denoted by $C_{l}$, can be obtained as

\begin{equation}\label{linkstaterate}
    C_{l}=\sum _{f=1}^{F}c_{f}(\alpha _{l}\hat{\textbf{C}}_{l}^{c_{f}}\textbf{1})
\end{equation}
where $\alpha _{l}$ denotes the steady-state probability that the
channel is in state $f$. Our highway mobility model is illustrated
in Fig. \ref{fig:mrsu2}, where a single V2R system with one coverage
region shown as a disk.

\begin{definition0}[\textbf{The probability of an OBU}]
Let $ D$ be the distance between a RSU and an OBU, and let $P_{f}$
be the probability of an OBU in link data rate $C_{f}$. According to
Figure \ref{fig:mrsu2}, $P_{f}$ is

\begin{equation} \label{eq:OBUprobability1}
P_{f}=\left\{
         \begin{aligned}
            P( r_{f+1}<D\leq r_{f}), &~~for~ f=1, 2, \cdots, N-1 \\
            P( 0<D\leq r_{f}),~~~~ &~~for~ f=N
         \end{aligned} \right.
\end{equation}
where $r_{f}$ is depicted in Figure \ref{fig:mrsu2}.
\end{definition0}

\begin{lemma0}[\textbf{The cumulative distribution function}]\label{lm:OBUcdf}
The cumulative distribution function of an OBU moving to the
position $(x,0)$ with length $d$ is

\begin{equation} \label{eq:OBUcdf}
F_{X}(x)=P(X\leq x)=\left\{
         \begin{aligned}
            &\frac{2x^{2}}{d^{2}}(\ln d -\ln x)+\frac{x^{2}}{d^{2}}, 0<x\leq d \\
            &~0, ~~~~~~~~~~~ x\leq 0 \\
            &~1, ~~~~~~~~~~~ x>d
         \end{aligned} \right.
\end{equation}

\end{lemma0}

The detailed proof of lemma \ref{lm:OBUcdf} can be seen in
\cite{ge2010optimal}. From (\ref{eq:OBUprobability1}) and
(\ref{eq:OBUcdf}), we obtain the probability of an OBU as follows:

%

\begin{multline}\label{eq:OBUprobability2}
P_{f}(f\neq N)=\frac{(1+2n\ln d)(s_{f}^{2}-s_{f+1}^{2})}{d^{2}}-\\
~~\frac{2s_{f}^{2}\ln s_{f}-2s_{f+1}^{2}\ln s_{f+1}}{d^{2}}+\frac{(1+2\ln d)(s_{2N-f}^{2})}{d^{2}}\\
-\frac{2s_{2N-f}^{2}\ln s_{2N-f}-2s_{2N-f+1}^{2}\ln
s_{2N-f+1}}{d^{2}}
\end{multline}

\begin{multline}\label{eq:OBUprobability3}
P_{f}(f= N)=\frac{(1+2n\ln d)(s_{N}^{2}-s_{N+1}^{2})}{d^{2}}\\
-\frac{2s_{N}^{2}\ln s_{N}-2s_{N+1}^{2}\ln s_{N+1}}{d^{2}}
\end{multline}

Let $l_{f}$ denote variable frame size in the region $f$, which
includes payload, MAC and physical layer header. According to
(\ref{eq:OBUprobability2}) and (\ref{eq:OBUprobability3}), the PGF
for frame size within the transmission range of the RSU is:

\begin{equation}\label{PGFofsuccesstransmissiontime}
S_{f}(z)=z^{rts+cts+3sifs}\sum
_{f=1}^{N}P_{f}z^{\frac{l_{f}+ACK}{C_{f}}}
\end{equation}

where sifs and ack is denoted as duration of the SIFS and ACK period
in slots, respectively. Additionally, $C_{f}$ is the average
transmission rate obtained by eq (\ref{linkstaterate}) and $P_{f}$
can be obtained from the expression (\ref{eq:OBUprobability2}) and
(\ref{eq:OBUprobability3}). In a RTS/CTS model, the PGF for the
collision period is given by $C_{f}(z)=z^{rts+cts+sifs}$.

\subsection{Distribution of Vehicles}~\label{OBUdistribution}
We now assume that $n$ OBUs move at a speed $V$ on a straight line
highway segment with length $d$. There are $n_{i}$ OBUs for type $i$
in the coverage area of the RSU ($i=1,2,\cdots,C$). Let $X_{n}^{i}$
be the distance between the n-th and the (n+1)-th vehicle of type i
that entered the AP¡¯s coverage area. In Table \ref{tab:notations},
we summarize the various quantities and notations we will use
throughout the paper.

\begin{table}[h]
\centering \caption{Summary of Notations} \label{tab:notations}
\begin{tabular}{|c|c|}
\hline Variable & Description\\
\hline $\lambda ^{i}$ & The average number of type i vehicles\\
\hline $x_{m}^{i}$ & A minimum inter-vehicle distance constraint\\
\hline q & The average number of vehicles per unit time\\
\hline $S(t)$ & The current back-off stage of the tagged node\\
\hline $B(t)$ & The back-off time of the tagged node\\
\hline $s^{i}_{m}$ & The minimum inter-OBU distance\\
\hline $\omega$ & The maximum number of OBUs\\
\hline $d$ & The maximum length of road segment\\
\hline $R_{jam}$ & The road capacity/length and $R_{jam}=d/\omega$\\
\hline $T_{o}$ & The transmission overhead in slots\\
\hline $T_{c}$ & The RTS collision overhead in slots\\
\hline $T_{s}$ & The payload transmission duration in slots\\
\hline
\end{tabular}
\end{table}

\begin{definition0}[\textbf{Renewal Process}]\label{df:renewalprocess}
Let $X_{n}^{i}$ and $N^{i}(d)$ represent the distance between the
$n$-th and the $(n+1)$-th OBU of type $i$, and the number of
vehicles of type $i$ over the length $d$ meters i.e. the RSU
coverage region, respectively. If the sequence of nonnegative random
variables $\{X_{1}^{i},X_{2}^{i},\cdots\}$ is independent and
identically distributed, then the counting process $\{N^{i}(d),
d\geq 0\}$ is is said to be a renewal process.
\end{definition0}

From Definition \ref{df:renewalprocess}, in vehicular traffic stream
models, $N^{i}(d)$ is a renewal process. Let
$D_{n}^{i}=\sum_{k=0}^{n} X_{n}^{i}$ denote the distance of the nth
renewal, $N^{i}(d)$ may be written as $N^{i}(d)=max\{n: D_{n}^{i}
\leq d\}$ \cite{ross2009introduction}. We obtain

\begin{equation}\label{eq:countDistribution}
    P\{N^{i}(d)=n\}=F_{n}^{i}(d)-F_{n+1}^{i}(d)
\end{equation}
where $F^{i}_{n}$ the n-fold convolution of $F^{i}$ with itself. It
is well known that the cumulative distribution function of
$X_{n}^{i}$,  the detailed $F^{i}$ can be seen in
\cite{tan2009modeling}. $F^{i}_{n}$ from
\cite{aydougdu2005pointwise,bruno2011throughput} may be written as

\begin{equation}\label{eq:convolution}
    F_{n}^{i}(d)=1-\sum_{j=0}^{n}\frac{(\lambda
    ^{i})^{j}}{j!}(d-nx_{m}^{i})^{n}e^{-\lambda ^{i}}(d-nx_{m}^{i})
\end{equation}

Let $\pi^{i}(n)$ denote the probability having n vehicles of type i
under the RSU coverage region. According to
(\ref{eq:countDistribution}) and (\ref{eq:convolution}), when $d=L$,
we have

\begin{multline}\label{eq:1}
\pi^{i}(n)=P\{N^{i}(d)=n\}=F_{n}^{i}(d)-F_{n+1}^{i}(d)\\
~~~~~~=\Sigma_{k=0}^{n}\frac{(\lambda ^{i})^{k}}{k!}[(\omega ^{i}-n-1)x^{i}_{m}]^{k}e^{-\lambda ^{i}[(\omega ^{i}-n-1)x^{i}_{m}]}\\
~~~~~~~-\Sigma_{k=0}^{n-1}\frac{(\lambda ^{i})^{k}}{k!}[(\omega ^{i}-n)x^{i}_{m}]^{k}e^{-\lambda ^{i}[(\omega ^{i}-n)x^{i}_{m}]}\\
n<\omega ^{i}-1
\end{multline}
\begin{equation}\label{eq:2}
\pi^{i}(n)=1-\Sigma_{k=0}^{n-1}\frac{(\lambda
^{i})^{k}}{k!}(x^{i}_{m})^{k}e^{-\lambda ^{i} x^{i}_{m}},n=\omega
^{i}-1 ~~
\end{equation}
\begin{equation}\label{eq:3}
\pi^{i}(n)=0, n\geq \omega
^{i}~~~~~~~~~~~~~~~~~~~~~~~~~~~~~~~~~~~~~~~~
\end{equation}

Furthermore, we define $\Gamma_{n}$ as $\Gamma_{n}=\{\textbf{n}:
\sum _{i=1}^{C}n_{i}=n, 0\leq n_{i}\leq \omega _{i}\}$, where
$\omega _{i}$ denote the maximum number for OBUs with type $i$ in a
single RSU coverage region and satisfies $d=\omega _{i}s^{i}_{m}$.
Additionally, $s^{i}_{m}$, $\omega$ and $d$ are defined in Table
\ref{tab:notations}. Thus, from \cite{bruno2011throughput}, the
probability that there are $n$ generic OBUs in a single RSU coverage
region is calculated as $\bar{\pi}_{n}=\sum_{\textbf{n}\in
\Gamma_{n}}\bar{\pi}(\textbf{n})=\prod_{i=1}^{C}\pi ^{i}(n_{i})$.
\subsection{Model of IEEE 802.11p Features}
Let us assume that all OBUs are identical, and analyzing the
behavior of one node make it enough to predict the behavior of the
other nodes and the channel performance. We denote this node as the
tagged node. Like \cite{felemban2011single}, the size of contention
window at backoff stage $j$, $W_{j}$, is defined as

\begin{equation} \label{eq:contentionwindow}
W_{_j }  = \left\{ {\begin{array}{*{20}c}
   {2^j CW_{min} } & {0 \le j \leq m}  \\
   {2^m CW_{min} } & {m < j < M}  \\
\end{array}} \right.
\end{equation}

where $CW_{min}$ denote the the minimum contention window size of
nodes and $m=\log_{2}(\frac{CW_{max}}{CW_{min}})$. $CW_{max}$ is the
maximum contention window size.
\subsection{Collision Probability}
In this section, we introduce the fixed-point equation detailed in
\cite{bianchi2000performance,kumar2007new}, which controls the
collision probability under saturation and non-saturation regimes.
Let $\beta^{c}$  denote the average attempt rate when the buffer is
not empty. The general attempt rate can be calculated by
$\beta=(1-p_{0})\beta^{c}$, where $p_{0}$ denote the probability
that the buffer is empty and computed in \cite{felemban2011single}.
By using the results in \cite{kumar2007new}, $\beta^{c}$ can be
given by:

\begin{equation}\label{collisionattemptrate}
\beta^{c}(\gamma)=\frac{\sum_{i=0}^{m}\gamma^{i}}{\sum_{i=0}^{m}b_{i}\gamma^{i}}
\end{equation}

Thus, by substituting $\beta^{c}$ and $p_{0}$ into the above
expression of $\beta$, the general collision probability $\gamma$
can be solved by the following fixed-point equation
$\gamma=\Gamma(\beta)$ ,

\begin{equation}\label{fixed-pointequation}
    \gamma=\Gamma((1-p_{0})\beta^{c})
\end{equation}

\subsection{Calculation for The Packet Service Time}\label{qwertasdf}
Assuming $T^{c}$ to be the service time distribution (in slots) of a
packet of a tagged node on the condition that the buffer is not
empty. Let $\chi$ be a random variable representing the time (in
slots) that elapses for one decrement of the back-off counter.

\begin{equation}\label{servicetimecount}
  T^{bf}=\sum _{i=1}^{T}\chi
\end{equation}
where $T=\Sigma _{j=0}^{k}b_{j}$, with the probability that the
packet transmission finishes at the kth back-off stage, $p(\gamma,
k), 0\leq k\leq M-1$, and $p(\gamma, k)$ is given by

\begin{equation}\label{contendprobability}
p(\gamma ,k) = \left\{ {\begin{array}{*{20}c}
   {(1 - \gamma )\gamma ^k } & {k = 0, \cdots m}  \\
   {\gamma ^{m + 1} } & {k = m + 1~~}  \\
\end{array}} \right.
\end{equation}

The generic slot duration $\chi$ depends on whether a slot is idle
or interrupted by a successful transmission or a collision. We
define $\chi$ as

\begin{equation}\label{slotduration}
    \chi = \left\{ {\begin{array}{*{20}c}
   {\begin{array}{*{20}c}
   {\sigma ,~~~~~~~~~~~} & {w.p. P_{i}}  \\
\end{array}}  \\
   {\begin{array}{*{20}c}
   {T_{s}+T_{o}+\sigma ,} & {w.p. P_{s}}  \\
\end{array}}  \\
{\begin{array}{*{20}c}
   {T_{c}+\sigma ,~~~~~~} & {w.p. P_{c}}  \\
\end{array}}  \\
\end{array}} \right.
 \end{equation}
where $P_{i}$ , $P_{s}$ and $P_{c}$ denote the steady state
probabilities of the channel being in idle, successful or collision
state, respectively, and ``w.p.'' means ``with the probability''.

Furthermore, we can introduce Channel State Markov Chain (CSMC)
defined in \cite{felemban2011single} to calculate the transition
probabilities $p_{ei}$, $p_{es}$, $p_{ec}$, $p_{si}$, $p_{ss}$,
$p_{ci}$, $p_{cs}$, and $p_{cc}$. Thus, $P_{i}$ , $P_{s}$ and
$P_{c}$ are calculate as follows:

\begin{equation}\label{slotduration}
\left( {\begin{array}{*{20}c}
   {p_{ei} } & {p_{es} } & {p_{ec} }  \\
   {p_{si} } & {p_{ss} } & 0  \\
   {p_{ci} } & {p_{cs} } & {p_{cc} }  \\
\end{array}} \right)\left( {\begin{array}{*{20}c}
   {P_i }  \\
   {P_s }  \\
   {P_c }  \\
\end{array}} \right) = \left( {\begin{array}{*{20}c}
   {P_i }  \\
   {P_s }  \\
   {P_c }  \\
\end{array}} \right)
 \end{equation}

Let $(b_{j})_{g}$, $\chi _{g}$, $T_{g}$, and $T_{g}^{ser}$ denote
the generating function of $b_{j}$, $\chi$, $T$ and $T^{ser}$
respectively, we have

 \begin{equation}\label{bpgf}
    (b_{j})_{g}(z) = \left\{ {\begin{array}{*{20}c}
   {\begin{array}{*{20}c}
   {\frac{1-z^{CW_{j}}}{CW_{j}(1-z)},~~~~~~~~} & {j=0, \cdots, m~~~~~}  \\
\end{array}}  \\
   {\begin{array}{*{20}c}
   {\frac{1-z^{CW_{m}}}{CW_{m}(1-z)},~~} & {j=m+1, \cdots, M-1}  \\
\end{array}}  \\
\end{array}} \right.
 \end{equation}

\begin{equation}\label{chipgf}
\chi
_{g}(z)=P_{i}z^{\sigma}+P_{s}z^{T_{s}+T_{o}+\sigma}+P_{c}z^{T_{c}+\sigma},
\end{equation}

\begin{equation}\label{Tpgf}
T_{g}(z)=\sum _{k=0}^{M-1}[p(\gamma, k)\prod _{j=0}^{k}(b_{j})_{g}]
\end{equation}

\begin{equation}\label{Tcollitionpgf}
T_{g}^{ser}(z)=\sum _{f=1}^{N}P_{f}T_{g}(\chi _{g}(z))
\end{equation}

Assume $S_{k}(T_{s})=kT_{s}$, and then $T_{s}$, $T_{c}$  and $T_{o}$
are defined in Table \ref{tab:notations}. In the basic access (BA)
case, the service time on the overall service requiring k attempts,
is given by:

if $0\leq k\leq m$,
\begin{equation*}
    T(k)= (k+1)T_{o}+T_{s}+S_{k}(T_{s})+\sum _{k=0}^{k}[p(\gamma, k)\prod _{j=0}^{k}(b_{j})_{g}]
\end{equation*}

if $ k= m+1$,
\begin{equation*}
    T(m+1)=(m+1)T_{o}+S_{m}(T_{s})+\sum _{k=0}^{m}[p(\gamma, k)\prod
                  _{j=0}^{k}(b_{j})_{g}]
\end{equation*}

In the RTS/CTS case, $T(k)$ and $T(m+1)$ is obtained by:

if $0\leq k\leq m$,
\begin{equation*}
    T(k)= kT_{c}+T_{o}+T_{s}+\sum _{k=0}^{k}[p(\gamma, k)\prod _{j=0}^{k}(b_{j})_{g}]
\end{equation*}

if $ k= m+1$,
\begin{equation*}
    T(m+1)=(m+1)T_{c}+\sum _{k=0}^{m}[p(\gamma, k)\prod
                  _{j=0}^{k}(b_{j})_{g}]
\end{equation*}

From eq (\ref{chipgf}), (\ref{Tpgf}) and (\ref{Tcollitionpgf}), the
Laplace transforms of the service time pdf, $L_{T}(s)$, in the BA
and RTS/CTS cases are, respectively, given by

\begin{equation}\label{qwe5}
\begin{split}
   L_{T}(s)=&\sum _{k=0}^{m}(1-\gamma)\gamma^{k}e^{-s(k+1)T_{o}}\sum _{l=1}^{M}q_{l}e^{-sa_{l}}\varphi _{k}(s)g_{l}^{k}(s) \\
    & +\gamma ^{m+1}e^{-s(m+1)T_{o}}\varphi _{m}(s)g_{l}^{m+1}(s)
\end{split}
\end{equation}
and
\begin{equation}\label{qwe5}
\begin{split}
   L_{T}(s)=&\sum _{k=0}^{m}(1-\gamma)\gamma^{k}E[e^{-sT(k)}]+\gamma ^{m+1}E[e^{-sT(m+1)}] \\
    =&\sum _{l=1}^{M}q_{l}e^{-sa_{l}} \sum
    _{k=0}^{m}(1-\gamma)\gamma^{k}e^{-s(T_{o}+kT_{c})}\varphi
    _{k}(s)\\
&+\gamma ^{m+1}e^{-s(m+1)T_{c}}\varphi _{m}(s)
\end{split}
\end{equation}

where $g_{l}(s)=\sum _{j=1}^{M}e^{-s \max \{a_{l},a_{j}\}}(\psi
_{j}-\psi _{j-1})$, $\psi _{j}=(1-\tau+\tau
Q_{j})^{n-1}/(1-(1-\tau)^{n-1})$ for $j=1,\cdots,M$ and
$\varphi_{k}(s)=\prod_{i=0}^{k}\frac{1-E[e^{-s\chi}]^{W_{i}}}{W_{i}[1-E[e^{-s\chi}]]}$.
By derivation of the above expression, the first order moment of
$T$, $E[T]|_{s=0}=(1-\gamma ^{m+1})E[U]/\theta(n)$ ($\theta(n)$ is
given in the following
section.)\cite{baiocchi2009variability,abate1992fourier}. According
to eq (\ref{eqn_dbl_y}), the total average service time is obtained
by:

\begin{equation}\label{serviceaverage}
\begin{split}
   & T_{ser} =\sum_{f=1}^{f=N}P_{f}(1-\gamma ^{m+1})n(1+[1-(
1-\beta)^{n}]T_{c}\\
    &+n(\beta
(1-\beta)^{n-1})(T_{o}-T_{c}) +n\beta (1-\beta)^{n-1}[\sum
_{i=1}^{n}\frac{1}{n}(\frac{L_{i}}{C_{f}})])
\end{split}
\end{equation}

\subsection{Throughput in The Same Channel}
Let us now consider a simpler situation where all OBUs are the
transmitter for a single flow and all packet lengths are equal to
$L$. The network throughput of the tagged OBU is given from
\cite{kumar2007new} at the top of the next page.

\newcounter{mytempeqncnt}
\begin{figure*}[!t]
\normalsize
\begin{equation}\label{eqn_dbl_y}
\theta (n)= \frac{L_{i}}{\frac{1}{\beta
(1-\beta)^{n-1}}+n(T_{o}-T_{c})+[\frac{1}{\beta
(1-\beta)^{n-1}}+(1-\frac{1}{\beta})]T_{c}+n[\sum
_{i=1}^{n}\frac{1}{n}(\frac{L_{i}}{\sum _{f=1}^{N}P_{f}C_{f,i}})]}
\end{equation}
\hrulefill
\vspace*{4pt}
\end{figure*}

%
\section{A Multichannel Population Game}~\label{MultichannelPopulationGame}
\subsection{Problem Formulation}
As the above mentioned, in this paper the focus of our consideration
is how to maximize the upload throughput and minimize service time
in V2R communications. To this end, we introduce a previously
mentioned population game \cite{sandholm2001potential}. Let $n_{c}$
denote the number of OBUs belonging to class $c$, we have $n=\sum
_{i=1}^{C}n_{i}$ and $n^{j}=\sum _{i=1}^{C}x^{j}_{i}$ to denote the
sum of all OBU's demands and the number of a single channel's active
OBUs respectively. From eq (\ref{eqn_dbl_y}), the throughput
received by the total mass of users of class $c$ connected to
channel $j$ is
\begin{equation}\label{classcmassthroughput}
\theta _{c}^{j}(\textbf{x}^{j})=
\frac{x^{j}_{c}L_{c}}{k_{0}^{j}+n^{j}[\sum_{f=1}^{N}P_{f}\sum
_{i=1}^{C}\frac{x^{j}_{i}}{n^{j}}(\frac{L_{i}}{C_{f,i}^{j}})]}
\end{equation}
\begin{equation*}
\begin{split}
k_{0}^{j}=&\frac{1}{\beta^{j}
(1-\beta^{j})^{n^{j}-1}}+n^{j}(T_{o}-T_{c})\\
&~~~~+[\frac{1}{\beta^{j}
(1-\beta^{j})^{n^{j}-1}}+(1-\frac{1}{\beta^{j}})]T_{c}
\end{split}
\end{equation*}

Then, we can obtain the throughput and service time per unit mass
respectively,
\begin{equation}\label{classcmassthroughput}
e _{c}^{j}(\textbf{x}^{j})=
\frac{L_{c}}{k_{0}^{j}+\sum_{f=1}^{N}P_{f}\sum
_{i=1}^{C}x^{j}_{i}(\frac{L_{i}}{C_{f,i}^{j}})}
\end{equation}
\begin{equation}\label{serviceaverage11}
\begin{split}
  T^{j}_{ser}= & (1-(\gamma^{j})^{m+1})(k_{0}^{j}\beta^{j}
(1-\beta^{j})^{n^{j}-1} \\
    & ~~~~+n^{j}\beta^{j}
(1-\beta^{j})^{n^{j}-1}[\sum_{f=1}^{N}P_{f}\sum
_{i=1}^{C}\frac{x^{j}_{i}}{n^{j}}(\frac{L_{i}}{C_{f,i}^{j}})])
\end{split}
\end{equation}
\begin{equation}\label{qwer11}
    T^{j}_{ser}=k_{1}^{j}+k_{2}^{j}\sum_{f=1}^{N}P_{f}\sum
_{i=1}^{C}\frac{x^{j}_{i}}{n^{j}}(\frac{L_{i}}{C_{f,i}^{j}})
\end{equation}
where $k_{1}^{j}=(1-(\gamma^{j})^{m+1})k_{0}^{j}\beta^{j}
(1-\beta^{j})^{n^{j}-1}$ and
$k_{2}^{j}=(1-(\gamma^{j})^{m+1})n^{j}\beta^{j}
(1-\beta^{j})^{n^{j}-1}$.

Thus, our problem is described in the following expression.
\begin{equation}\label{maxminfunction:po}
\begin{split}
   \max _{\textbf{x}}&(\frac{\sum_{n=1}^{\omega}\overline{\pi}_{n}}{1-\pi_{0}}\sum _{i=1}^{C}\pi^{i}(n_{i})\sum _{j=1}^{L}x^{j}_{i}e_{i}^{j}(\textbf{x})\\
       & -\frac{\sum_{n=1}^{\omega}\overline{\pi}_{n}}{1-\pi_{0}}\sum
   _{i=1}^{C}\zeta ^{i}\pi^{i}(n_{i})\sum
   _{j=1}^{L}x^{j}_{i}T_{ser}(x^{j}_{i}))
   \end{split}
\end{equation}
subject to $\sum _{j=1}^{L}x^{j}_{i}=n_{i}, \forall i\in \{1,
\cdots, C\}$;$x_{^{(i)}}^{j}\ge 0$ and $x^{j}_{i}=0$ if channel $j$
is not provided to OBUs of class $i$.

To obtain the optimal solution satisfying the expression, let us see
the following Lemma \ref{lm:throughputservicetime}. In order to
facilitate the description, we let
$\Phi=\frac{\sum_{n=1}^{\omega}\overline{\pi}_{n}}{1-\pi_{0}}$. The
calculations of the following variables, are given in Section
\ref{OBUdistribution}.

\begin{lemma0}\label{lm:throughputservicetime}
A game $F$ potential function can be given by
\begin{equation}\label{gamefunction}
   \begin{split}
     \Theta (\textbf{x})= & \Phi\sum _{i=1}^{C}\pi^{i}(n_{i})\sum _{j=1}^{L}x^{j}_{i}e_{i}^{j}(\textbf{x})\\
       & -\Phi\sum
   _{i=1}^{C}\zeta ^{i}\pi^{i}(n_{i})\sum _{j=1}^{L}x^{j}_{i}T_{ser}(x^{j}_{i})
   \end{split}
\end{equation}
with $x^{j}_{i}=0$ and $\zeta$ is a weight that provides influence
to service time versus the throughput for the tagged OBU, if channel
$j$ is not available to $i$ class OBUs.
\end{lemma0}
\emph{Proof:} The proof of the lemma can be found in Appendix.

Obviously, the above results to meet Definition
\ref{df:potentialgame}. Thus, The function $\Theta (\textbf{x})$ is
a potential function for the game $F$.
\penalty10000\hfill\penalty10000\vrule height 5pt depth 2pt width
7pt \penalty-10000

From the above proof, we obtain the payoff function per unit mass
for OBUs of class $c$ in channel $l$,
$F_{c}^{l}=\frac{\sum_{n=1}^{\omega}\overline{\pi}_{n}}{1-\pi_{0}}(\pi^{c}(n_{c})e
_{c}^{l}(\textbf{x})-\wp _{c}^{l}(\textbf{x}) \ast \sum
_{i=1}^{C}\pi^{i}(n_{i})\theta_{i}^{l}(\textbf{x})
-k^{l}_{2}\frac{L_{c}}{n^{l}\bar{C}^{l}}\sum
_{i=1}^{C}\zeta_{i}\pi_{i}(n_{i})x_{i}^{l}-\zeta_{c}\pi_{c}(n_{c})T^{l}_{ser})$.

\begin{lemma0}\label{lm:Wardropequilibrium}
$\textbf{x}^{*}=(x^{(1)*}, \cdots, x^{(C)*})$ is called a Wardrop
equilibrium if the payoff function per unit mass for OBUs of class
$c$ in channel $l'$ of the non-cooperative game $F$ is given by

\begin{equation}\label{payofffunction}
\begin{split}
F_{c}^{l'}(\textbf{x})=&\frac{\sum_{n=1}^{\omega}\overline{\pi}_{n}}{1-\pi_{0}}(\pi^{c}(n_{c})e
_{c}^{l'}(\textbf{x})-\wp _{c}^{l'}(\textbf{x}) \ast \sum
_{i=1}^{C}\pi^{i}(n_{i})\theta_{i}^{l'}(\textbf{x})\\
&~~~~-k^{l'}_{2}\frac{L_{c}}{n^{l'}\bar{C}^{l'}}\sum
_{i=1}^{C}\zeta_{i}\pi_{i}(n_{i})x_{i}^{l'}-\zeta_{c}\pi_{c}(n_{c})T^{l'}_{ser}),
\end{split}
\end{equation}
where for each $c$ we have
\begin{equation*}
    \begin{array}{*{20}c}
   {(x_{^{(c)}}^{j})^{*}\ge 0,\forall j,c}&{\sum\limits_{j =1}^{L} {(x_{^{(c)}}^{j} )^{*}= n_{c},\forall c,}}&{F_{c}^{l'}(\textbf{x}^{*})\ge F_{c}^{l} (\textbf{x}^{*})}\\
\end{array}
\end{equation*}
\end{lemma0}
\emph{Proof:} From Definition (\ref{df:wardropequilibrium}), it
holds evidently. \penalty10000\hfill\penalty10000\vrule height 5pt
depth 2pt width 7pt \penalty-10000

Further, it is obvious that all obtained vectors $\textbf{x}$ have
equal payoffs if eq (\ref{payofffunction}) take the zero value in
all channels $l$. As such, a Wardrop equilibrium is obtained. Take
the previous mentioned BNN dynamics into account, we have the
following Lemma.
\begin{lemma0}\label{lm:throughputservicetimeKTGM}
The potential game $F$ equilibrium satisfies the equation
(\ref{maxminfunction:po})
\end{lemma0}

%
%
\emph{Proof:} For Definition (\ref{df:bnn}), we know BNN dynamics
and its PC \cite{shakkottai2006case}. Thus,
$F_{c}^{l}(\textbf{x})=\frac{1}{n^{c}}\sum _{j\in \mathcal
{S}}x_{c}^{j}F_{c}^{j}(\textbf{x})$ or $x_{c}^{l}=0$.  Then we can
construct a optimization problem as follows:

\begin{equation}\label{optimizationconstruct}
 \begin{split}
    \min _{\eta}\max _{\textbf{x}}(&\frac{\sum_{n=1}^{\omega}\overline{\pi}_{n}}{1-\pi_{0}}\sum _{i=1}^{C}\pi^{i}(n_{i})\sum _{j=1}^{L}x^{j}_{i}e_{i}^{j}(\textbf{x})\\
       & ~~-\frac{\sum_{n=1}^{\omega}\overline{\pi}_{n}}{1-\pi_{0}}\sum
   _{i=1}^{C}\zeta ^{i}\pi^{i}(n_{i})\sum
   _{j=1}^{L}x^{j}_{i}T_{ser}(x^{j}_{i})\\
   &~~~~-\sum _{i=1}^{C}\eta _{i}(\sum _{j=1}^{L}x^{j}_{i}-n_{i}))
    \end{split}
\end{equation}

the solution of $\eta _{i}$ satisfies $\eta
_{i}=F_{i}(\textbf{x})=\frac{1}{n^{i}}\sum _{j\in \mathcal
{S}}x_{i}^{j}F_{i}^{j}(\textbf{x})$. Therefore, the solutin of the
potential game $F$ satisfies the expression
(\ref{optimizationconstruct}). We denote it as $\Theta
(\textbf{x}^{*})$. Since the expression (\ref{maxminfunction:po}) is
not concave, the solutions of the expression
(\ref{optimizationconstruct}) and (\ref{maxminfunction:po}) are
equal. \penalty10000\hfill\penalty10000\vrule height 5pt depth 2pt
width 7pt \penalty-10000

From Lemma \ref{lm:throughputservicetime},
\ref{lm:Wardropequilibrium}, we can know that selfish OBUs can
attain the maximum throughput and the minimum service time by
selecting one of those channels hold $F_{c}^{l'}(\textbf{x})=0$.
Lemma \ref{lm:throughputservicetimeKTGM} shows that our solution
satisfying a Wardrop equilibrium condition also is the optimal
solution.
\subsection{A Game Among the OBUs and RSU}
We now define the price vector as $\textbf{p}=<p_{1},\cdots,p_{L}>$,
which is the price the RSU provides for different type OBUs.
Contrary to \cite{shakkottai2006case}, our policy is oriented for
different type OBUs, instead of different channels. The reason that
our policy is reasonable is that high-speed moving type OBUs should
be different from low-speed OBUs like walkers.

The RSU expects to provide the bandwidth for the types with the
maximum price, so its gain function $\Psi=\Phi\sum _{j=1}^{L}\sum
_{i=1}^{C}\zeta ^{i}\pi^{i}(n_{i})x^{j}_{i}p_{i}$. However, the OBUs
want to give the minimum price to meet their demands. Thus, a game
exists among the OBUs and RSU. Consequently, they converge to the
equilibrium of the system and the corresponding to potential
function is defined as follows:

\begin{lemma0}\label{lm:throughputservicetimegame}
A game $F$ potential function can be given by
\begin{equation}\label{gamefunction}
   \begin{split}
&\Theta (\textbf{x})=  \Phi\sum _{i=1}^{C}\pi^{i}(n_{i})\sum _{j=1}^{L}x^{j}_{i}e_{i}^{j}(\textbf{x})\\
        &-\Phi\sum
   _{i=1}^{C}\zeta ^{i}\pi^{i}(n_{i})\sum
   _{j=1}^{L}x^{j}_{i}T_{ser}(x^{j}_{i})-\Phi\sum
   _{j=1}^{L}\sum
   _{i=1}^{C}\zeta ^{i}\pi^{i}(n_{i})x^{j}_{i}p_{i}
   \end{split}
\end{equation}
with $x^{j}_{i}=0$ if channel $j$ is not available to $i$ class
OBUs.
\end{lemma0}

\emph{Proof:}Certification process is similar to that in Lemma
\ref{lm:throughputservicetime}.
\penalty10000\hfill\penalty10000\vrule height 5pt depth 2pt width
7pt \penalty-10000

Further, taking the previous mentioned BNN dynamics into account, we
have the following Lemma.

\begin{lemma0}\label{lm:throughputservicetimeKTGMgame}
The potential game $F$ equilibrium can be calculated by the
following equation
\begin{equation}\label{maxminfunction}
\begin{split}
   &\max _{\textbf{x}}(\Phi\sum _{i=1}^{C}\pi^{i}(n_{i})\sum _{j=1}^{L}x^{j}_{i}e_{i}^{j}(\textbf{x})\\
   &-\Phi\sum
   _{i=1}^{C}\zeta ^{i}\pi^{i}(n_{i})\sum
   _{j=1}^{L}x^{j}_{i}T_{ser}(x^{j}_{i})-\Phi\sum
   _{j=1}^{L}\sum
   _{i=1}^{C}\zeta ^{i}\pi^{i}(n_{i})x^{j}_{i}p_{i}
   \end{split}
\end{equation}
subject to
\begin{equation*} \sum _{j=1}^{L}x^{j}_{i}=n_{i}, \forall
i\in \{1, \cdots, C\}; x_{^{(i)}}^{j}\ge 0
\end{equation*}
and $x^{j}_{i}=0$ if channel $j$ is not provided to OBUs of class
$i$.
\end{lemma0}

\emph{Proof:}Certification process is similar to that in Lemma
\ref{lm:throughputservicetimeKTGM}.
\penalty10000\hfill\penalty10000\vrule height 5pt depth 2pt width
7pt \penalty-10000

From Lemma \ref{lm:throughputservicetimegame} and
\ref{lm:throughputservicetimeKTGMgame}, the RSU can attain the
optimal performance by selecting the price vector that makes the
maximum throughput and minimum service time.
\section{Case Study and Simulations}~\label{CaseStudy}
In this section, we test our channel scheduling performance for our
uplink data access and validate successfully with the NS-2
simulator. In our scenario, there are three wireless channels, i.e.
802.11a, 802.11b and 802.11g, and a highway model of length 1.2Km,
with two lanes. all OBUs compete for the channel applying IEEE
802.11 DCF. The values for the parameters of vehicular velocity and
delay are illustrated in Table \ref{tab:notations}.

\begin{table}[h]
\centering \caption{The velocity and deadline parameters of OBUs}
\vspace{5pt} \label{tab:notations}
\begin{tabular}{|c|c|}
\hline Variable & Value\\
\hline $v_{max}$(m/s) & 35\\
\hline  $v_{min}$(m/s) & 10\\
\hline an OBU deadline belonging to type 1(ms) & 0.2\\
\hline an OBU deadline belonging to type 2(ms) &0.35\\
\hline
\end{tabular}
\end{table}

According to the parameters from Table \ref{tab:notations}, firstly,
we use VanetMobiSim simulator to produce a TCL script about
vehicular mobility. Secondly we add a fixed RSU to the TCL scipt.
Thirdly, we explore the vehicle density impact on throughput and the
number of data transferred respectively by using the modified script
as the input of NS2. Finally, according to experimental results, we
make analysis about our optimal policy.

\begin{figure}[H]
\vspace{10pt} \centering
\begin{minipage}{200pt}
\centering
    \includegraphics[width=0.75 \textwidth]{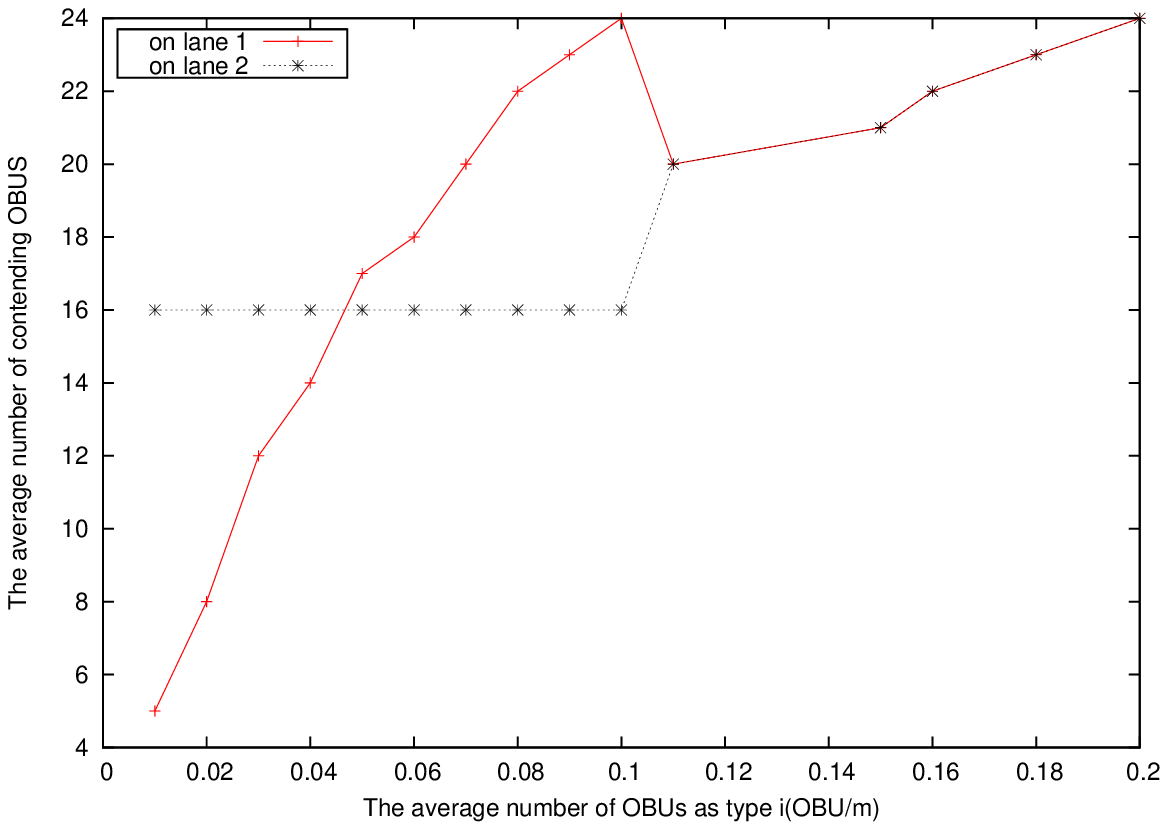}
    \caption{Average number of contending vehicles versus $\lambda ^{1}(\lambda
^{2}=0.03)$.} \label{fig:price1}
\end{minipage}
\hspace {10pt}%
\begin{minipage}{200pt}
\centering
    \includegraphics[width=0.75
    \textwidth]{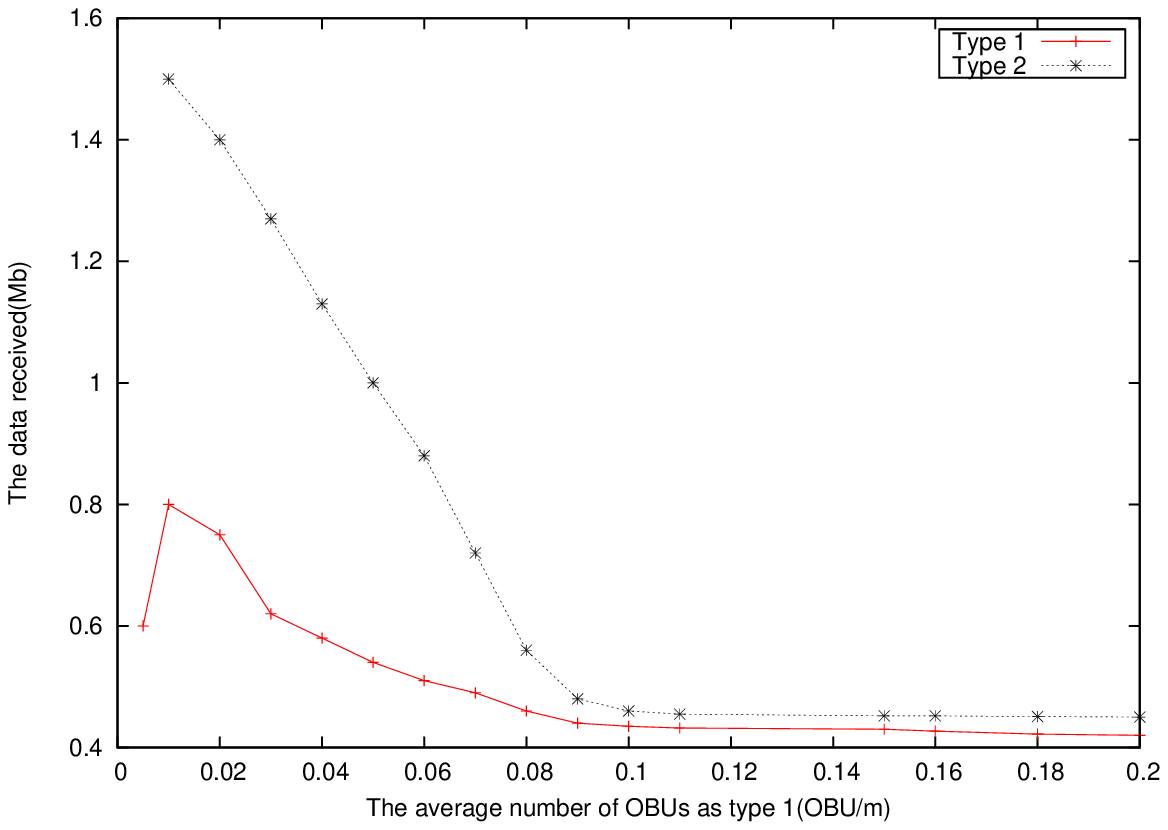}
    \caption{The number of data transferred by the OBU from different types under our equilibrium policy versus$\lambda
^{1}(\lambda ^{2}=0.03)$.} \label{fig:price2}
\end{minipage}
\end{figure}

In Fig.~\ref{fig:price1}, we fix the density of OBU type 2 as
$\lambda ^{2}=0.03$ and assume  the density of OBU type 1 $\lambda
^{1}(\lambda ^{2}=0.03)$ changes from $0$ to $2\rho$, to which is
the results of $\lambda ^{2}$ are similar. This is the result of a
vehicle simulation scenarios construction, and further research is
beyond the scope of our discussion. Fig.~\ref{fig:price1} shows our
model predictions are accurate.

\begin{figure*}
\vspace{10pt} \centering \subfigure[]{ \label{fig:randoma}
\includegraphics[width=2.2in]{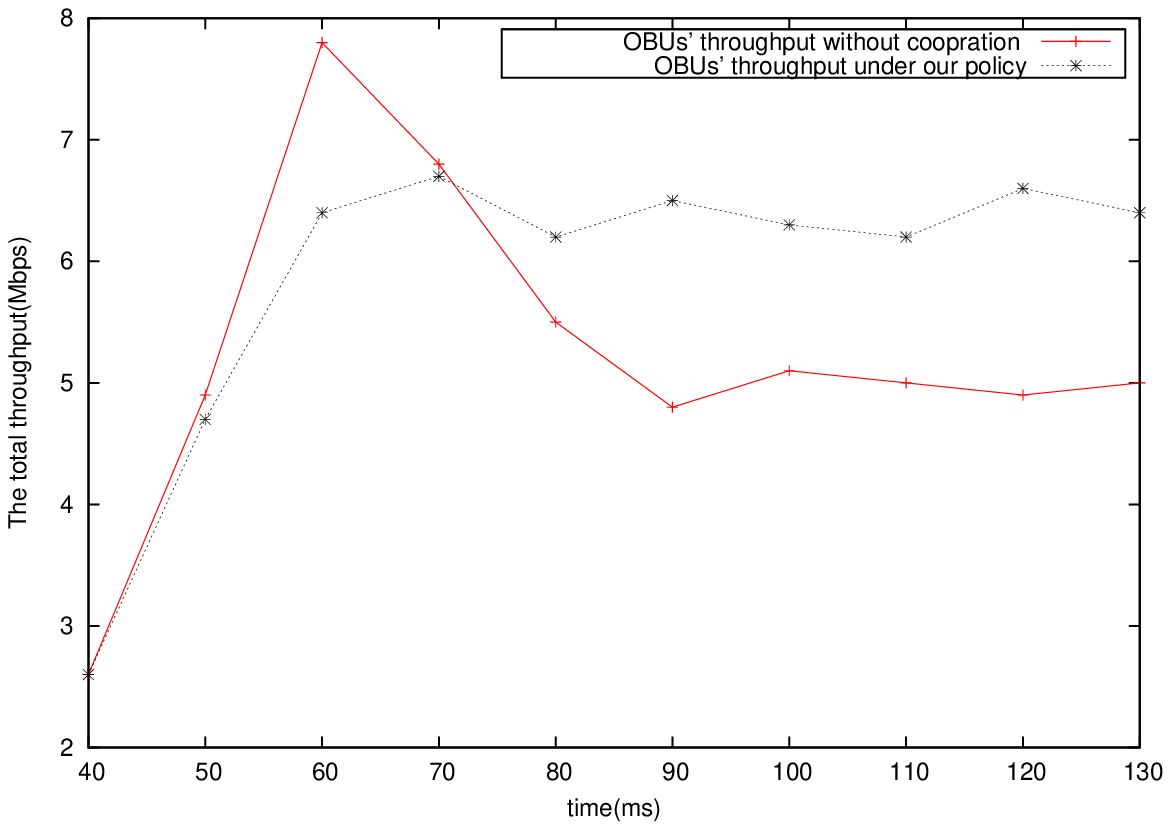}}
\hspace{-0.13in} \subfigure[]{ \label{fig:randomb}
\includegraphics[width=2.2in]{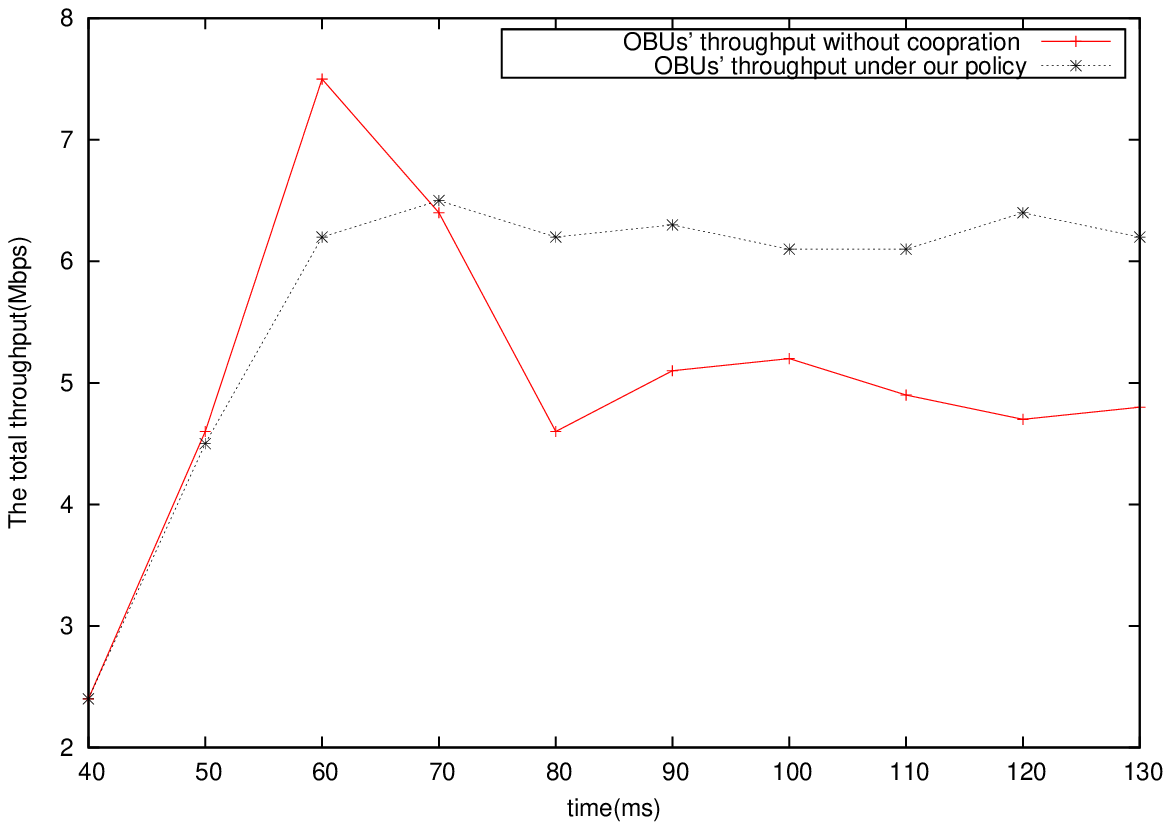}}
\hspace{-0.13in} \subfigure[]{ \label{fig:randomc}
\includegraphics[width=2.2in]{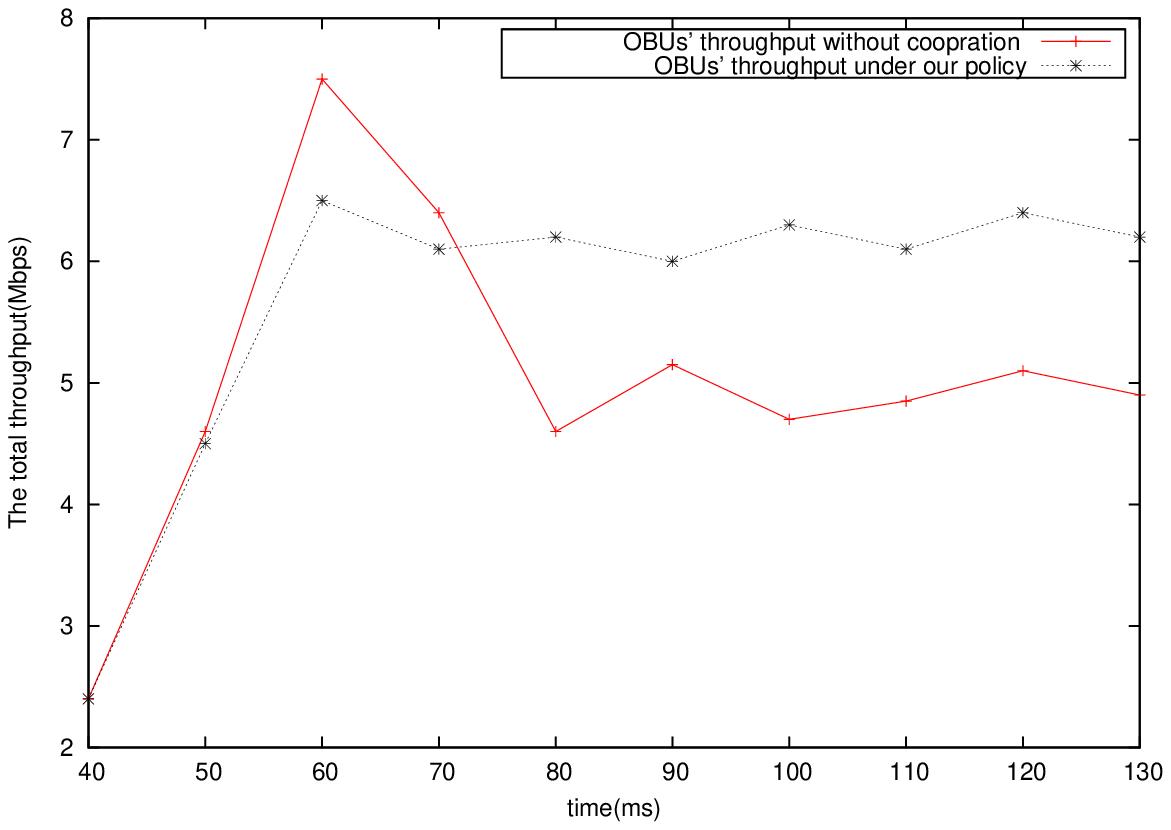}}
\caption{The throughput versus time in fixed $\lambda ^{2}=0.03$,
$\lambda ^{1}=0.05$, $\lambda ^{1}=0.1$ and $\lambda ^{1}=0.2$. (a)
$\lambda ^{1}=0.05(\lambda ^{2}=0.03)$; (b) $\lambda
^{1}=0.1(\lambda ^{2}=0.03)$;(c) $\lambda ^{1}=0.2(\lambda
^{2}=0.03)$.} \label{fig:random} \vspace{10pt}
\end{figure*}

\begin{figure*}
\centering \subfigure[]{ \label{fig:randomaa}
\includegraphics[width=2.2in]{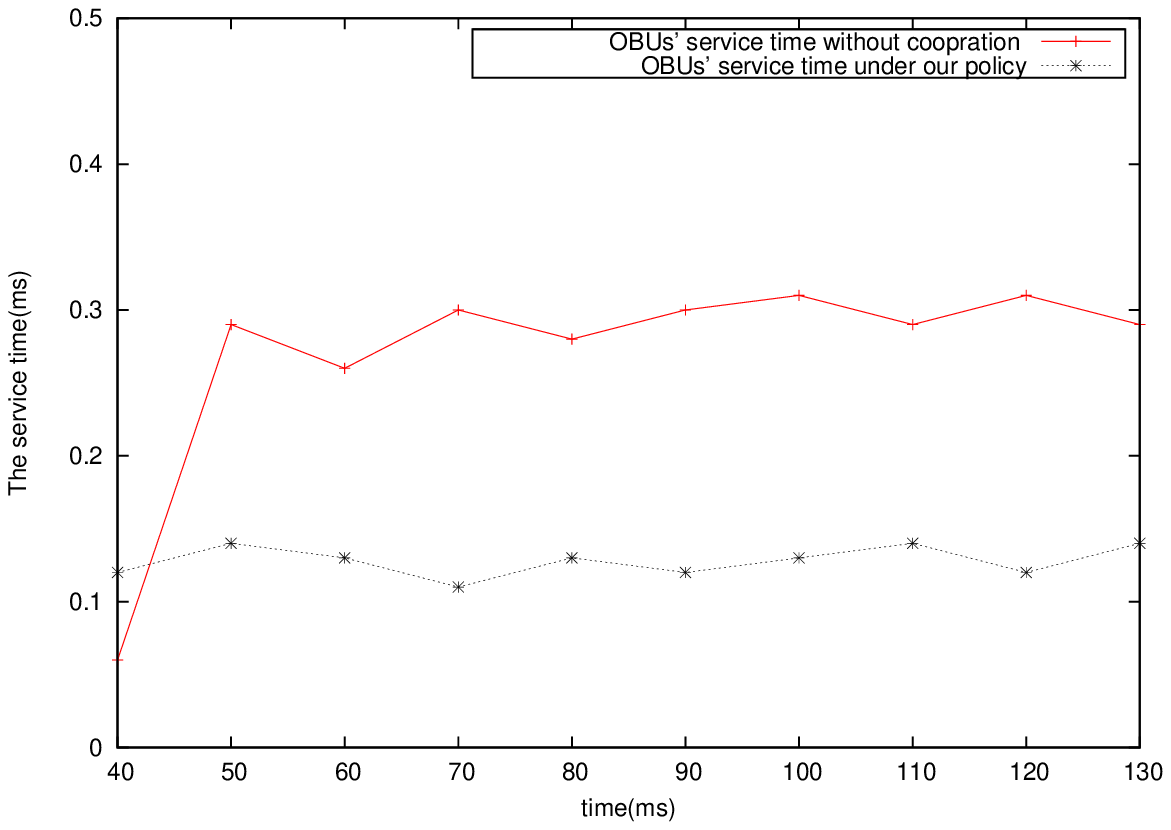}}
\hspace{-0.15in} \subfigure[]{ \label{fig:randombb}
\includegraphics[width=2.2in]{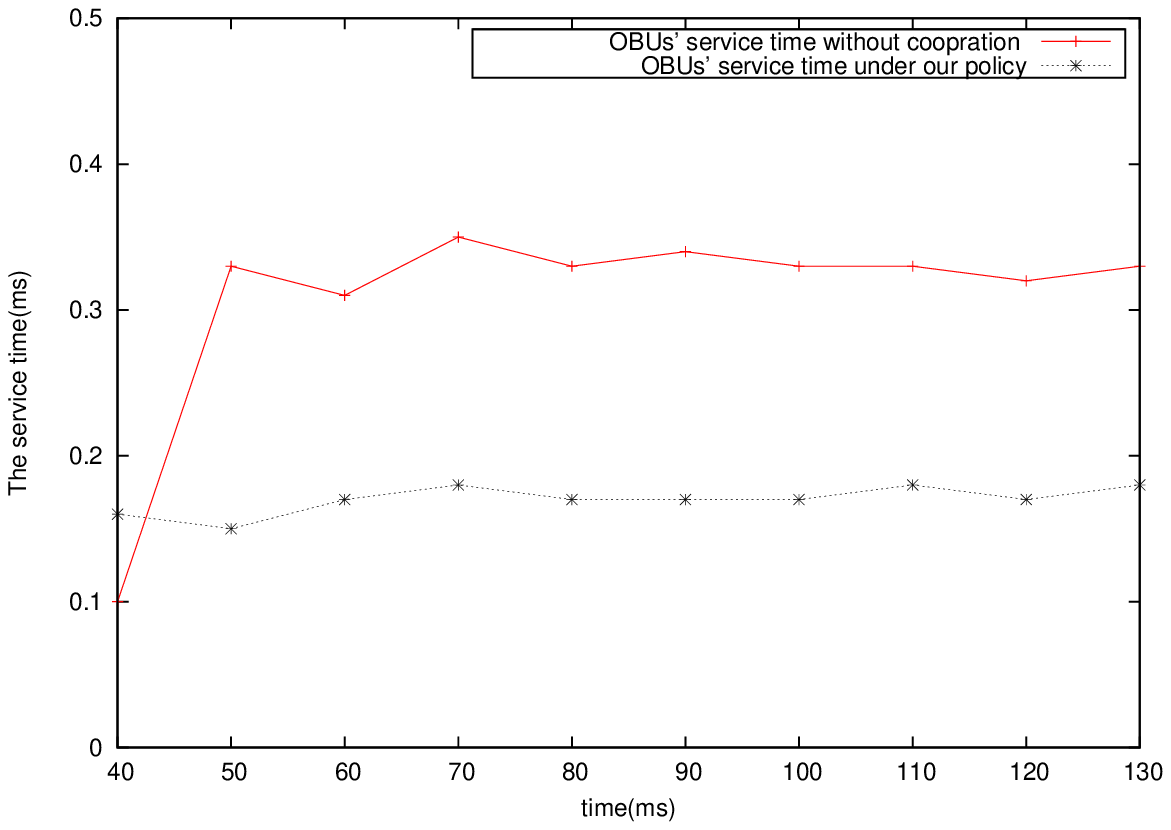}}
\hspace{-0.15in} \subfigure[]{ \label{fig:randomcc}
\includegraphics[width=2.2in]{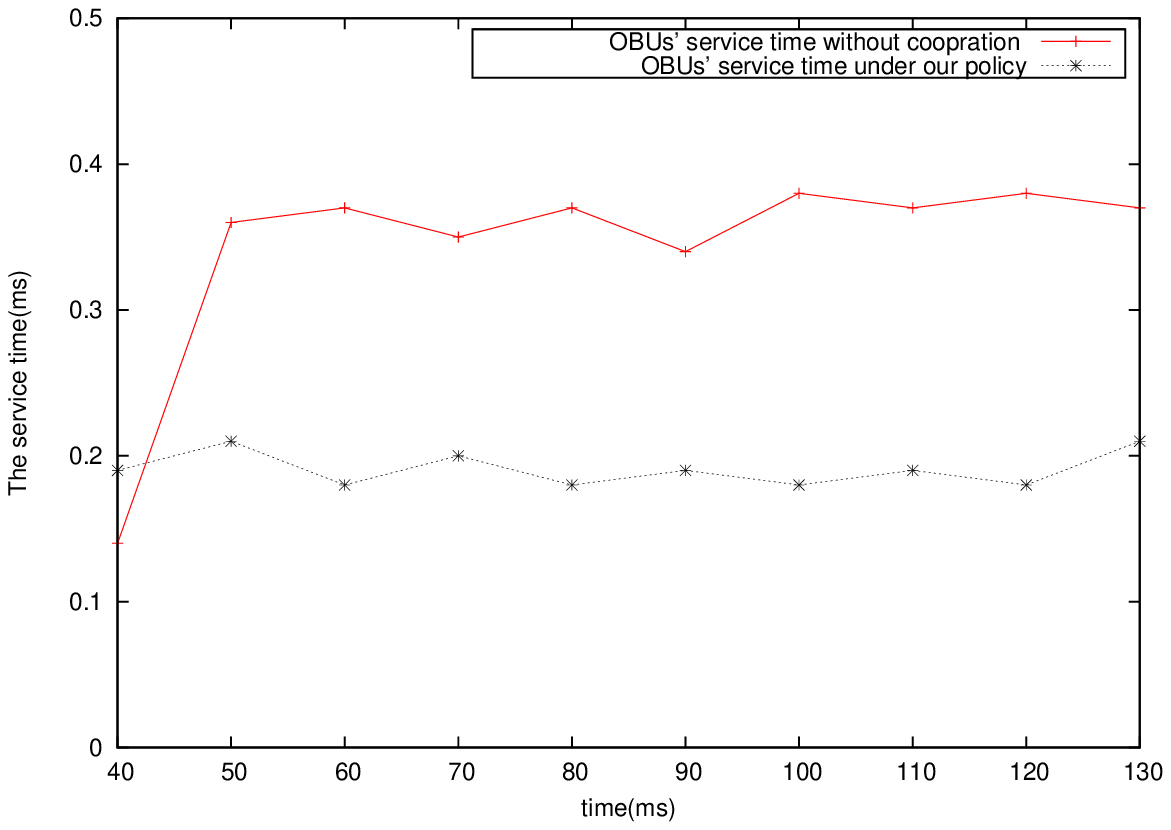}}
\caption{The service time of data packets versus time in $\lambda
^{2}=0.03$, $\lambda ^{1}=0.05$, $\lambda ^{1}=0.1$ and $\lambda
^{1}=0.2$. (a) $\lambda ^{1}=0.05(\lambda ^{2}=0.03)$; (b) $\lambda
^{1}=0.1(\lambda ^{2}=0.03)$; (c) $\lambda ^{1}=0.2(\lambda
^{2}=0.03)$.} \label{fig:randomq} \vspace{3pt}
\end{figure*}

Fig.~\ref{fig:price2} demonstrates that the number of data
transferred by the OBU from different types under our equilibrium
policy change with $\lambda ^{1}(\lambda ^{2}=0.03)$. First, we can
observe the amount of data transmitted is not a monotone decreasing
function of the average number of OBUs per unit distance $\rho$.
Second, under a VSN scenario, the number of packets transferred from
different types OBUs reveals the unfairness of a certain degree
(e.g. OBU type 1 and OBU type 2). this is because in the RSU
coverage area, slow OBUs have a longer length of stay and less
stringent delay requirements. However, the idea of the tradeoff
between the fairness and the parameters seem not very intuitive,
which makes it possible that our strategy can be optimized in the
service time.

We verify the correctness of the expression (\ref{serviceaverage})
and the expression (\ref{eqn_dbl_y}) in two group experiments. one
group is to verify the correctness of the network throughput
expression, the other group for the verification of the service
time. In particular, we have observed that there is a deviation in
the throughput performance, it is because an OBU through the RSU
coverage region needs some during to reach the steady state of a
backoff process. It is such a transition time caused the overall
throughput performance deviation. The experiments of both groups
have demonstrated the network throughput and service time calculated
from the expressions are very similar to the simulated environment
results based on the case for the distribution of
Fig.~\ref{fig:price2}. Further, our optimization policy based on
throughput performance and service time are compared with
\cite{misic2011performance}.

In Fig.~\ref{fig:random}, we test the throughput of OBUs versus time
in fixed $\lambda ^{2}=0.03$, $\lambda ^{1}=0.05$, $\lambda
^{1}=0.1$ and $\lambda ^{1}=0.2$ respectively. Experimental results
show that under the three density of OBUs, our Wardrop equilibrium
policy has better steady performance than the algorithm in
\cite{misic2011performance}. We also observed that as the vehicle
density increases, our policy has greater total throughput than
\cite{misic2011performance}.

In Fig.~\ref{fig:randomq}, we test the service time of data packets
from OBUs versus time in fixed $\lambda ^{2}=0.03$, $\lambda
^{1}=0.05$, $\lambda ^{1}=0.1$ and $\lambda ^{1}=0.2$ respectively.
Experimental results show that under the three density of OBUs, our
Wardrop equilibrium policy makes the service time of the data
packets to attain an earlier steady state than the algorithm in
\cite{misic2011performance}. We also observed that as the vehicle
density increases, the service time of the data packets has a
slighter increase. Even in the increase case, our policy also can
ensure inclusive service to basically meet delay requirements of
different type OBUs.

\section{Conclusions and Future Work}~\label{conclude}
With the development of wireless technology and popularity of
roadside multi-channel WiFi devices, more and more passengers in the
vehicle expect the enjoyment of high-bandwidth data transmissions
from the multi-channel wireless devices. So such high data
throughput and low latency scheduling problem for car passengers are
common concerns of industrial and academic fields. This paper
presents our system model and related definitions, and performs a
brief discussion of the non-cooperative games and population game.
Then, we analyze the actual link state communication model of OBUs
and a RSU. A single OBU type in a mathematical model is extended to
multiple types for OBUs. Further, we will combine the actually link
status and a Markov multi-type model based on different regions. The
throughput and service time expressions are further developed by
applying the collision probability. Finally, we have formed a
non-cooperative game problem. Theoretically we proved that the
solution of the balance point and the optimization problem is the
same. Further simulations also show that the solution of the
equilibrium point meets the requirements of  the maximum throughput
and service time. What's more, in future work, in order to protect
the car users to enjoy multi-hop scenario, high-bandwidth data
transmission, we will further study the timeliness issues of data
transmission and scheduling in a multi-hop scene.

\bibliographystyle{IEEEtran}

\bibliography{IEEEtran}
\appendix
\section* {Proof of Lemma \ref{lm:throughputservicetime}}

A game $F$ potential function can be given by
\begin{equation*}
   \begin{split}
     \Theta (\textbf{x})= & \Phi\sum _{i=1}^{C}\pi^{i}(n_{i})\sum _{j=1}^{L}x^{j}_{i}e_{i}^{j}(\textbf{x})\\
       & -\Phi\sum
   _{i=1}^{C}\zeta ^{i}\pi^{i}(n_{i})\sum _{j=1}^{L}x^{j}_{i}T_{ser}(x^{j}_{i})
   \end{split}
\end{equation*}
with $x^{j}_{i}=0$ and $\zeta$ is a weight that provides influence
to service time versus the throughput for the tagged OBU, if channel
$j$ is not available to $i$ class OBUs.

\emph{\textbf{Proof:}} From eq (\ref{classcmassthroughput}) and
(\ref{qwer11}), we obtain
\begin{equation*}
    \begin{split}
       \frac{\partial \Theta (\textbf{x})}{\partial x_{c}^{l}}=&\frac{\partial}{\partial x_{c}^{l}}[\Phi\sum _{i=1}^{C}\pi^{i}(n_{i})\sum _{j=1}^{L}x^{j}_{i}e_{i}^{j}(\textbf{x})] \\
         &-\frac{\partial}{\partial x_{c}^{l}}[\Phi\sum
   _{i=1}^{C}\zeta ^{i}\pi^{i}(n_{i})\sum
   _{j=1}^{L}x^{j}_{i}T_{ser}(x^{j}_{i})]
     \end{split}
\end{equation*}
where
\begin{equation*}
\begin{split}
    \frac{\partial}{\partial x_{c}^{l}}&[\Phi\sum _{i=1}^{C}\pi^{i}(n_{i})\sum _{j=1}^{L}x^{j}_{i}e_{i}^{j}(\textbf{x})]\\
    =&\frac{\partial}{\partial x_{c}^{l}}[\Phi\sum _{i=1}^{C}\pi^{i}(n_{i})\sum _{j=1}^{L}[\frac{x^{j}_{i}L_{i}}{k_{0}^{j}+\sum_{f=1}^{N}P_{f}\sum _{i=1}^{C}x^{j}_{i}(\frac{L_{i}}{C_{f,i}^{j}})}]]\\
=&\Phi[\frac{\pi^{c}(n_{c})L_{c}}{k_{0}^{l}+\sum_{f=1}^{N}P_{f}\sum
_{i=1}^{C}x^{l}_{i}(\frac{L_{i}}{C_{f,i}^{l}})}\\
&~~~-\frac{\sum_{f=1}^{N}P_{f}\frac{L_{c}}{C_{f,c}^{l}}\sum
_{i=1}^{C}\pi^{i}(n_{i})x^{l}_{i}L_{i}}{(k_{0}^{l}+\sum_{f=1}^{N}P_{f}\sum
_{i=1}^{C}x^{l}_{i}(\frac{L_{i}}{C_{f,i}^{l}}))^{2}}]\\
=&\Phi[\pi^{c}(n_{c})e
_{c}^{l}(\textbf{x})-[\frac{\sum_{f=1}^{N}P_{f}\frac{L_{c}}{C_{f,c}^{l}}}{k_{0}^{l}+\sum_{f=1}^{N}P_{f}\sum
_{i=1}^{C}x^{l}_{i}(\frac{L_{i}}{C_{f,i}^{l}})}\ast\\
&~~~\sum
_{i=1}^{C}\frac{\pi^{i}(n_{i})x_{i}^{l}L_{i}}{k_{0}^{l}+\sum_{f=1}^{N}P_{f}\sum
_{i=1}^{C}x^{l}_{i}(\frac{L_{i}}{C_{f,i}^{l}})}]]
\end{split}
\end{equation*}
where this occupancy factor of per unit mass is $\wp
_{c}^{l}(\textbf{x})=\frac{\sum_{f=1}^{N}P_{f}\frac{L_{c}}{C_{c}^{l}}}{k_{0}^{l}+\sum_{f=1}^{N}P_{f}\sum
_{i=1}^{C}x^{l}_{i}(\frac{L_{i}}{C_{f,i}^{l}})}=\frac{\frac{L_{c}}{\bar{C}^{l}}}{k_{0}^{l}+\sum_{f=1}^{N}P_{f}\sum
_{i=1}^{C}x^{l}_{i}(\frac{L_{i}}{C_{f,i}^{l}})}$. In fact, if we let
$C_{c}^{l}(\textbf{x})=\wp _{c}^{l}(\textbf{x}) \ast \sum
_{i=1}^{C}\pi_{i}(n_{i})\theta_{i}^{l}(\textbf{x})$ in the above
expression. From the RSU¡¯s aspect, $C_{c}^{l}(\textbf{x})$ denotes
the cost of a unit mass of OBUs of class $l$. Further,
\begin{equation*}
\begin{split}
    \frac{\partial}{\partial x_{c}^{l}}&[\Phi\sum
   _{i=1}^{C}\zeta ^{i}\pi^{i}(n_{i})\sum _{j=1}^{L}x^{j}_{i}T_{ser}(x^{j}_{i})]\\
    &=\frac{\partial}{\partial x_{c}^{l}}[\Phi\sum
   _{i=1}^{C}\zeta ^{i}\pi^{i}(n_{i})\\
   &~~~~~~~\sum _{j=1}^{L}x^{j}_{i}[k_{1}^{j}+k_{2}^{j}\sum_{f=1}^{N}P_{f}\sum
_{i=1}^{C}\frac{x^{j}_{i}}{n^{j}}(\frac{L_{i}}{C_{f,i}^{j}})]]\\
    &=\Phi(k_{2}^{l}\sum_{f=1}^{N}P_{f}\frac{L_{c}}{n^{l}C_{f,c}^{l}}\sum
_{i=1}^{C}\zeta_{i}\pi_{i}(n_{i})x_{i}^{l}\\
&~~~~~~~+\zeta_{c}\pi_{c}(n_{c})(k_{1}^{l}+k^{l}_{2}\sum_{f=1}^{N}P_{f}\sum_{i=1}^{C}\frac{x_{i}^{l}L_{i}}{n^{l}C_{f,i}^{l}}))\\
&=\Phi(k^{l}_{2}\sum_{f=1}^{N}P_{f}\frac{L_{c}}{n^{l}C_{f,c}^{l}}\sum
_{i=1}^{C}\zeta_{i}\pi_{i}(n_{i})x_{i}^{l}+\zeta_{c}\pi_{c}(n_{c})T^{l}_{ser})\\
&=\Phi k^{l}_{2}\frac{L_{c}}{n^{l}\bar{C}^{l}}\sum
_{i=1}^{C}\zeta_{i}\pi_{i}(n_{i})x_{i}^{l}+\Phi\zeta_{c}\pi_{c}(n_{c})T^{l}_{ser}
\end{split}
\end{equation*}
Therefore,
\begin{equation*}
    \begin{split}
\frac{\partial \Theta (\textbf{x})}{\partial
x_{c}^{l}}&=\Phi(\pi^{c}(n_{c})e _{c}^{l}(\textbf{x})\\
&~~~~~~-\wp _{c}^{l}(\textbf{x}) \ast \sum
_{i=1}^{C}\frac{\pi^{i}(n_{i})x_{i}^{l}L_{i}}{k_{0}^{l}+\sum_{f=1}^{N}P_{f}\sum
_{i=1}^{C}x^{l}_{i}(\frac{L_{i}}{C_{f,i}^{l}})}\\
&~~~~~~~~~~~-k^{l}_{2}\frac{L_{c}}{n^{l}\bar{C}^{l}}\sum
_{i=1}^{C}\zeta_{i}\pi_{i}(n_{i})x_{i}^{l}-\zeta_{c}\pi_{c}(n_{c})T^{l}_{ser})\\
=&\frac{\sum_{n=1}^{\omega}\overline{\pi}_{n}}{1-\pi_{0}}(\pi^{c}(n_{c})e
_{c}^{l}(\textbf{x})-\wp _{c}^{l}(\textbf{x}) \ast \sum
_{i=1}^{C}\pi^{i}(n_{i})\theta_{i}^{l}(\textbf{x})\\
&~~~-k^{l}_{2}\frac{L_{c}}{n^{l}\bar{C}^{l}}\sum
_{i=1}^{C}\zeta_{i}\pi_{i}(n_{i})x_{i}^{l}-\zeta_{c}\pi_{c}(n_{c})T^{l}_{ser})\\
=&F_{c}^{l}(\textbf{x})
\end{split}
\end{equation*}

Obviously, the above results to meet Definition
\ref{df:potentialgame}. Thus, The function $\Theta (\textbf{x})$ is
a potential function for the game $F$.
\penalty10000\hfill\penalty10000\vrule height 5pt depth 2pt width
7pt \penalty-10000
\end{document}